\documentclass[final]{nst}

\usepackage{subfigure,dcolumn}
\usepackage[T2A,T1]{fontenc}
\usepackage[russian,english]{babel}
\usepackage{siunitx}

\usepackage{listings}
\usepackage{tikz,xcolor,hyperref}
\definecolor{lime}{HTML}{A6CE39}
\DeclareRobustCommand{\orcidicon}{
\begin{tikzpicture}
\draw[lime, fill=lime] (0,0)
circle[radius=0.16]
node[white]{{\fontfamily{qag}\selectfont \tiny \.{I}D}}; 
\end{tikzpicture}
\hspace{-2mm}
}
\foreach \x in {A, ..., Z}{
\expandafter\xdef\csname orcid\x\endcsname{\noexpand\href{https://orcid.org/\csname orcidauthor\x\endcsname}{\noexpand\orcidicon}}
}

\begin{document}

\title{Preparation and measurement of an $\rm ^{37}$Ar source for liquid xenon detector calibration}
\thanks{This work was supported by National Key R\&D grant from the Ministry of Science and Technology of China (Nos. 2021YFA1601600, 2023YFA1606200),  National Science Foundation of China (Nos. 12090062, 12105008), and the Major State Basic Research Development Program of China.}

\author{Xu-Nan Guo\orcidA}
\affiliation{School of Physics, Beihang University, Beijing 102206, China}

\author{Chang Cai}
\affiliation{Department of Physics\&Center for High Energy Physics, Tsinghua University, Beijing 100084, China}

\author{Fei Gao}
\affiliation{Department of Physics\&Center for High Energy Physics, Tsinghua University, Beijing 100084, China}

\author{Yang Lei}
\affiliation{Department of Physics\&Center for High Energy Physics, Tsinghua University, Beijing 100084, China}

\author{Kai-Hang Li}
\affiliation{Department of Physics\&Center for High Energy Physics, Tsinghua University, Beijing 100084, China}

\author{Chun-Lei Su}
\affiliation{Northwest Institute of Nuclear Technology, Xi'an 710024, China}

\author{Ze-Peng Wu}
\affiliation{Northwest Institute of Nuclear Technology, Xi'an 710024, China}

\author{Xiang Xiao}
\affiliation{School of Physics, Sun Yat-Sen University, Guangzhou 510275, China}

\author{Ling-Feng Xie}
\email[Corresponding author, ]{xlf22@mails.tsinghua.edu.cn}
\affiliation{Department of Physics\&Center for High Energy Physics, Tsinghua University, Beijing 100084, China}

\author{Yi-Fei Zhao}
\affiliation{Department of Physics\&Center for High Energy Physics, Tsinghua University, Beijing 100084, China}

\author{Xiao-Peng Zhou\orcidB}
\email[Corresponding author, ]{xpzhou@buaa.edu.cn}
\affiliation{School of Physics, Beihang University, Beijing 102206, China}

\begin{abstract}

We present the preparation and measurement of the radioactive isotope $\rm ^{37}Ar$, which was produced using thermal neutrons from a reactor, as a calibration source for liquid xenon time projection chambers. $\rm ^{37}Ar$ is a low-energy calibration source with a half-life of 35.01 days, making it suitable for calibration in the low-energy region of liquid xenon dark-matter experiments. Radioactive isotope $\rm ^{37}Ar$ was produced by irradiating $\rm ^{36}Ar$ with thermal neutrons. It was subsequently measured in a gaseous xenon time projection chamber (GXe TPC) to validate its radioactivity. Our results demonstrate that $\rm ^{37}Ar$ is an effective and viable calibration source that offers precise calibration capabilities in the low-energy domain of xenon-based detectors.

\end{abstract}

\keywords{$\rm ^{37}Ar$, Gaseous Xenon detector, Low-energy, Calibration source}

\maketitle

\section{Introduction}
Xenon is an exceptional medium for particle detection due to its high density, large atomic mass, and excellent scintillation properties. The dual-phase xenon time projection chamber leverages the superior properties of xenon and is extensively utilized in dark matter\cite{pandax4t_dm,xenonnt_dm,lz_dm,pandax_er,sbdm,karl} searches, neutrino detection\cite{pandax_b8,xenonnt_b8,pandax_0vbb,xenon1t_0vbb,RELICS}, and related experiments. 
It is primarily based on the precise reconstruction of scintillation signals (S1) and ionization signals (S2) generated by particles that deposit energy in liquid xenon (LXe).
Scintillation photons, detected by photomultiplier tubes (PMTs), generate a pulse signal referred to as S1. 
The ionization electrons, under the influence of an extraction electric field, drift into the gaseous xenon phase and emit secondary scintillation light through the electroluminescence process, and are then recorded as S2.
The spatial coordinates of an event were reconstructed from the patterns of S1 and S2, with photoelectron counts proportional to the energy magnitude of the signal.
The geometric variation and inhomogeneous distribution of the electric field and light collection efficiency influence the detector and lead to a significant position dependence of the signal intensities of S1 and S2, which not only reduces the precision of the energy of events and three-dimensional position reconstruction but also weakens the ability to distinguish between nuclear and electronic recoil events\cite{3dmodel}.
Therefore, it is essential to use a calibration source that can be uniformly distributed in LXe and yield monoenergetic signals to calibrate the detector response.

Owing to its uniform mixing properties with xenon, the $\rm ^{37}Ar$ gaseous source has emerged as an ideal calibration source. 
The radioactive isotope $\rm ^{37}Ar$, with a half-life of 35.01 days, can decay to $\rm ^{37}Cl$ and neutrinos\cite{bipm-5} via the electron capture process.
During this process, the atomic nucleus captures an electron from the K, L, or M shell. 
The resulting vacancies were filled by outer electrons, accompanied by the emission of X-rays or Auger electrons.
The total energy deposition of these processes corresponds to the binding energyies of each shell: 2.82 keV (K-shell), 0.27 keV (L-shell), and 0.01 keV (M-shell), with decay branch ratios of 90.2\%, 8.7\%, and 1.1\%, respectively\cite{Renier,Akimov,Chechev,Aprile}. 
The energy depositions of the K and L shells were close to the energy threshold of the LXe dark matter detectors, making $\rm ^{37}Ar$ an ideal calibration source.
Furthermore, $\rm ^{37}Ar$ can be removed using a cryogenic distillation tower similar to that of $\rm ^{85}Kr$\cite{removekr}, thereby improving its potential application in detector calibration.

The production of $\rm ^{37}Ar$ has long been a subject of interest owing to its potential applications in various fields, including low-background detection and fundamental nuclear research. 
In the atmosphere, the primary source of $\rm ^{37}Ar$ is the reaction of fast neutrons produced by cosmic rays, $\rm ^{40}Ar$(n,4n)$\rm ^{37}Ar$\cite{ar40}. Although $\rm ^{40}Ar$ constitutes up to 99.60\% of natural argon, the cross-sectional effects result in a low yield of $\rm ^{37}Ar$, accompanied by the production of numerous other radioactive isotopes, particularly long-lived $\rm ^{39}Ar$, which is highly undesirable.
Another method for producing $\rm ^{37}Ar$ involves irradiating $\rm ^{40}Ca$ in calcium oxide (CaO) with fast neutrons\cite{Kelly}. This approach has been commonly used in the past owing to its high yield\cite{Ca40}. However, to facilitate the extraction of $\rm ^{37}Ar$ from CaO, the target material must be prepared in powdered form. Additionally, $\rm ^{37}Ar$ gas was subsequently distilled at high temperatures in a sealed container. This high-temperature distillation process imposes stringent requirements on the technology and equipment involved. Moreover, powdered CaO may be carried along with gas into the xenon detector, causing contamination. Impurities such as radon, which is co-distilled with $\rm ^{37}Ar$, can also interfere with low-background experiments.
Thermal neutron irradiation of $\rm ^{36}Ar$ is an effective technique for preparing radioactive isotopes $\rm ^{37}Ar$.
Although the reaction cross-section for $\rm ^{36}Ar$(n,$\gamma$)$\rm ^{37}Ar$ is lower than that for $\rm ^{40}Ca$(n,$\alpha$)$\rm ^{37}Ar$, the preparation of the target material is simpler, and the range of products is more limited. This method is particularly suitable for high-sensitivity, low-background experiments such as those used for dark matter detection.

We performed a detailed simulation based on Geant4 to identify the various nuclei expected to be produced after irradiation. 
In particular, considering the complexity of the energy distribution of the reactor neutron source, we need to avoid producing by-products such as $\rm ^{39}Ar$ which would produce a low-energy electronic recoil background in large-scale LXe detectors and would be difficult to remove.
This is because $\rm ^{37} Ar$ gas can be distributed in gaseous xenon at room temperature. 
We adopted the GXe TPC to measure $\rm ^{37}Ar$ radioactivity.

The remainder of this paper is organized as follows. Section~\ref{sec:setup} describes in detail the preparation of $\rm ^{37}Ar$, including the simulation and feasibility assessment. Section~\ref{sec:mearsurement} shows the measurement results of the activity of $\rm ^{37}Ar$ obtained through the operation and analysis of the gaseous xenon detector.

\section{Preparation of $\rm \mathbf{^{37}Ar}$ calibration source}
\label{sec:setup}

\subsection{Experimental Setup and Principles} 
\label{sec:exp-setup}

\begin{figure}[!htb] 
\includegraphics[width=\linewidth]{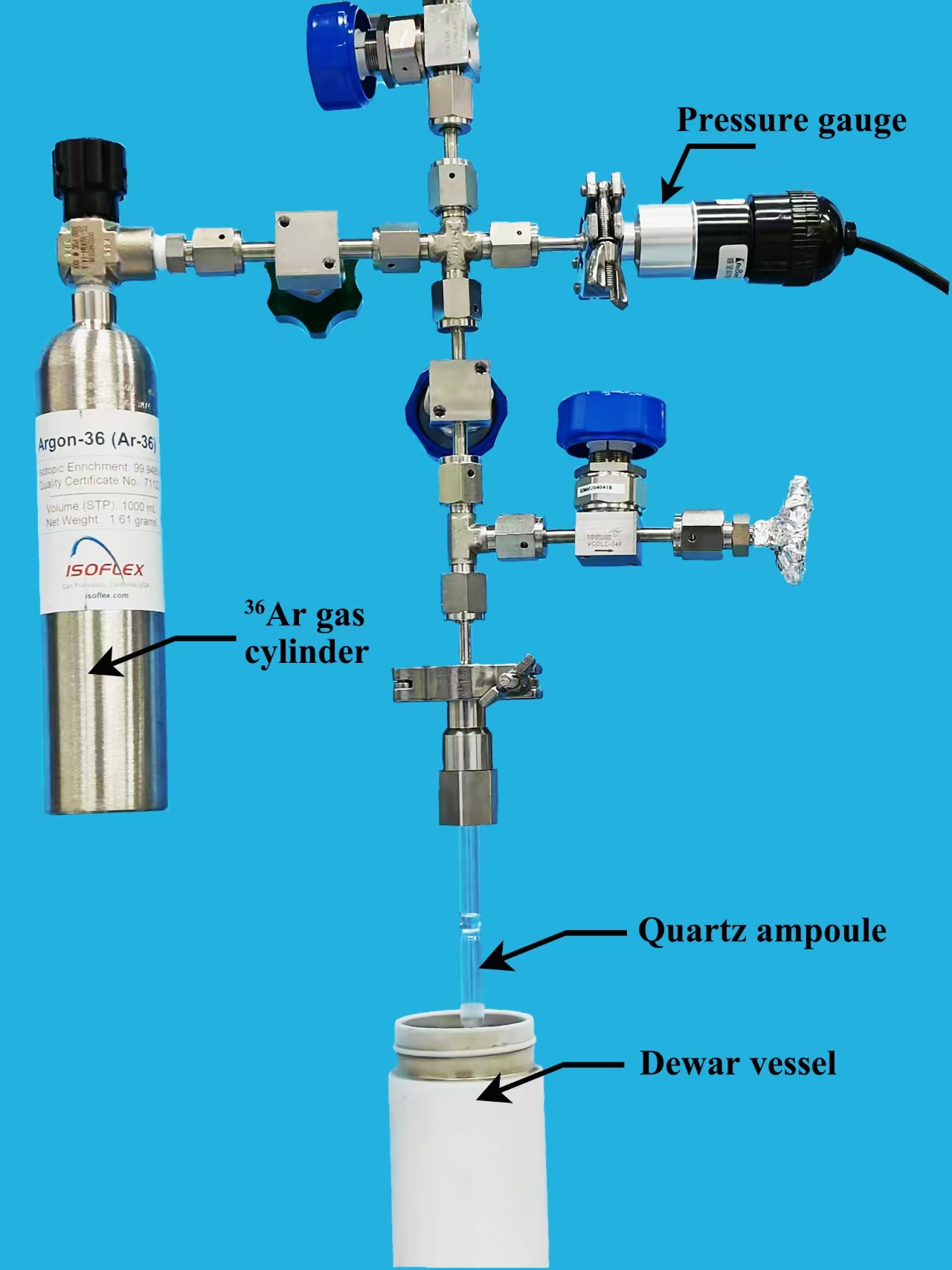}
\caption{(Color online) Melt-seal apparatus of quartz ampoule, incorporating an $\rm ^{36}Ar$ gas cylinder, a resistive silicon pressure gauge, and a dewar vessel filled with liquid nitrogen at the base.}
\label{fig:preparation_system}
\end{figure}

\begin{figure}[!htb] 
\subfigure{
\label{fig:anpule1}
\includegraphics[width=\hsize]{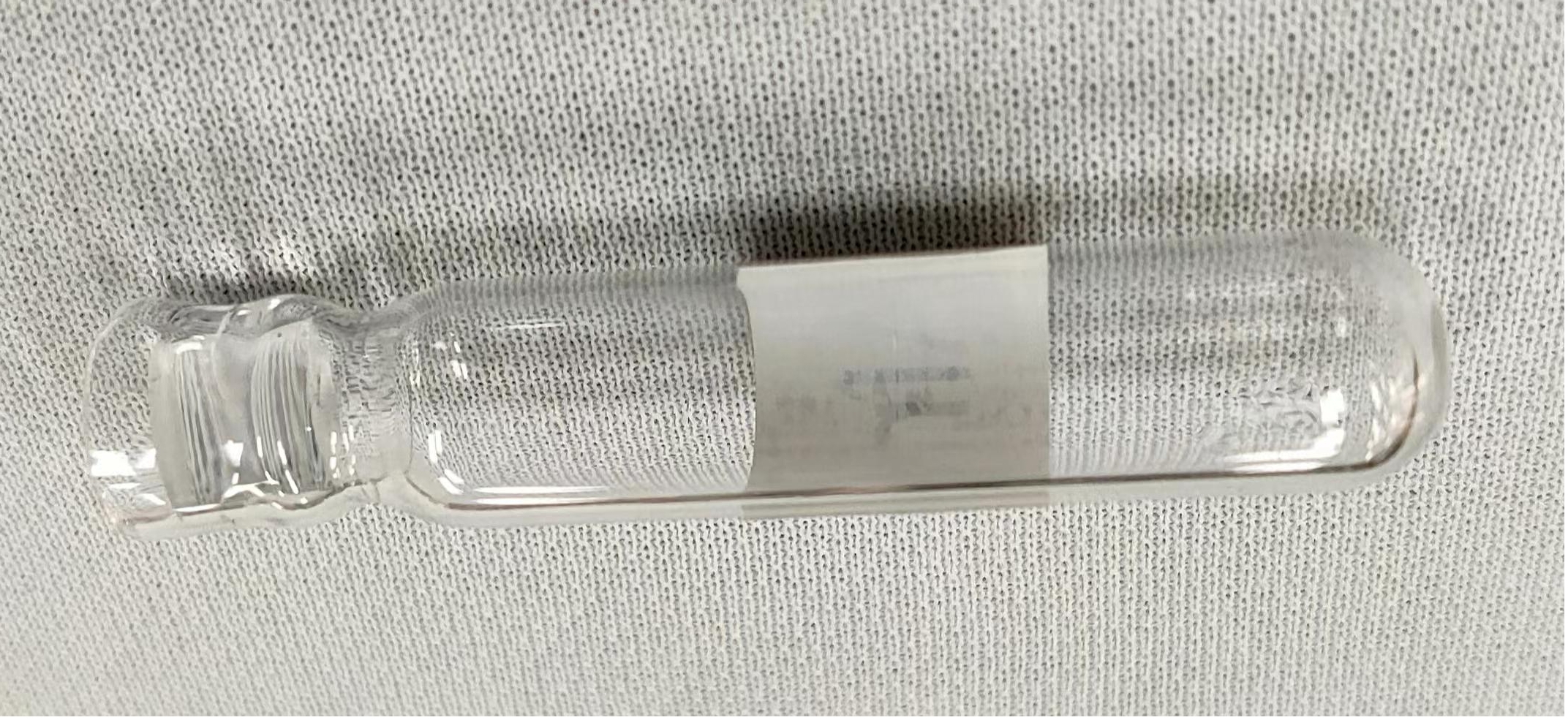}
}
\subfigure{
\label{fig:ampule2}
\includegraphics[width=\hsize]{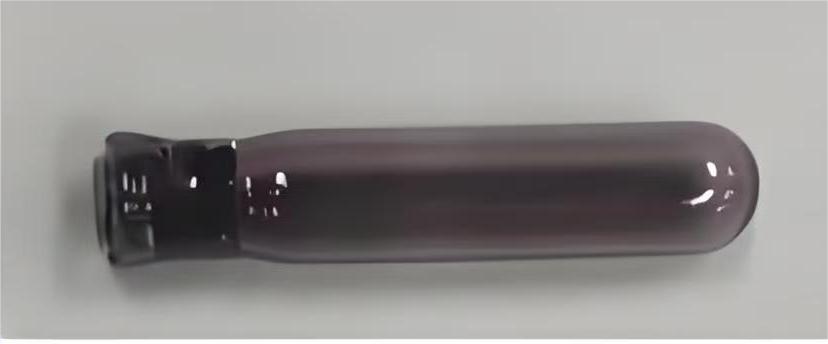}
}
\caption{(Color online) The quartz glass container (top) before irradiation and (bottom) after irradiation, with the wall thickness is 1 mm and the inner pressure is 0.4 bar.}
\label{fig:ampule} 
\end{figure}

The target isotope $\rm ^{37}Ar$ was produced by irradiating high-purity (99.935\%) $\rm ^{36}Ar$ with thermal neutrons. 
This process involved sealing $\rm ^{36}Ar$ in a precisely specified quartz ampoule with a diameter of 1 cm, length of 4 cm, and wall thickness of 1 mm. 
The relative pressure of the package was negative.
$\rm ^{37}Ar$ is produced by the neutrons captured by $\rm ^{36}Ar$.
The reactor neutron source\cite{Liu} generated a thermal neutron flux of $1.5 \times 10^{13}\ \mathrm{n/(cm^2 \cdot s)}$ with an irradiation duration of 2.17 hours.
Additionally, because of the intrinsic properties of the neutron source, an accompanying epithermal neutron flux of 6.25×10$^{11}$n/cm$^{2}$/s was present. 
The uncertainty in the neutron flux measurements was estimated at $5\%$.
Sealing of the quartz ampoule was a critical step in the experiment. A melt-seal technique was used in this process, as illustrated in Fig. ~\ref{fig:preparation_system}, we used liquid nitrogen on the bottom side of the quartz ampoule to create a low-temperature environment for enrichment of $\rm ^{36}Ar$. The other side was sealed with a high-temperature hydrogen torch. This method ensures the airtightness and structural integrity of the seal.
Figure~\ref{fig:ampule} shows the quartz ampule in its pre- and post-neutron irradiation states. The transformation of the ampule to a dark purple color was hypothesized to be the result of microscopic structural and chemical alterations induced by irradiation. Neutron irradiation catalyses the formation of color centers within the silicon dioxide matrix. These color centers introduce new energy levels within the electron bandgap, which leads to photothermal absorption. The superposition of various absorption bands results in the creation of absorption maxima, which in turn impart a tinting effect on the vitreous material \cite{Luo,Fu}.

\begin{figure}[!htb] 
\subfigure{
\label{fig:presure1}
\includegraphics[width=\hsize]{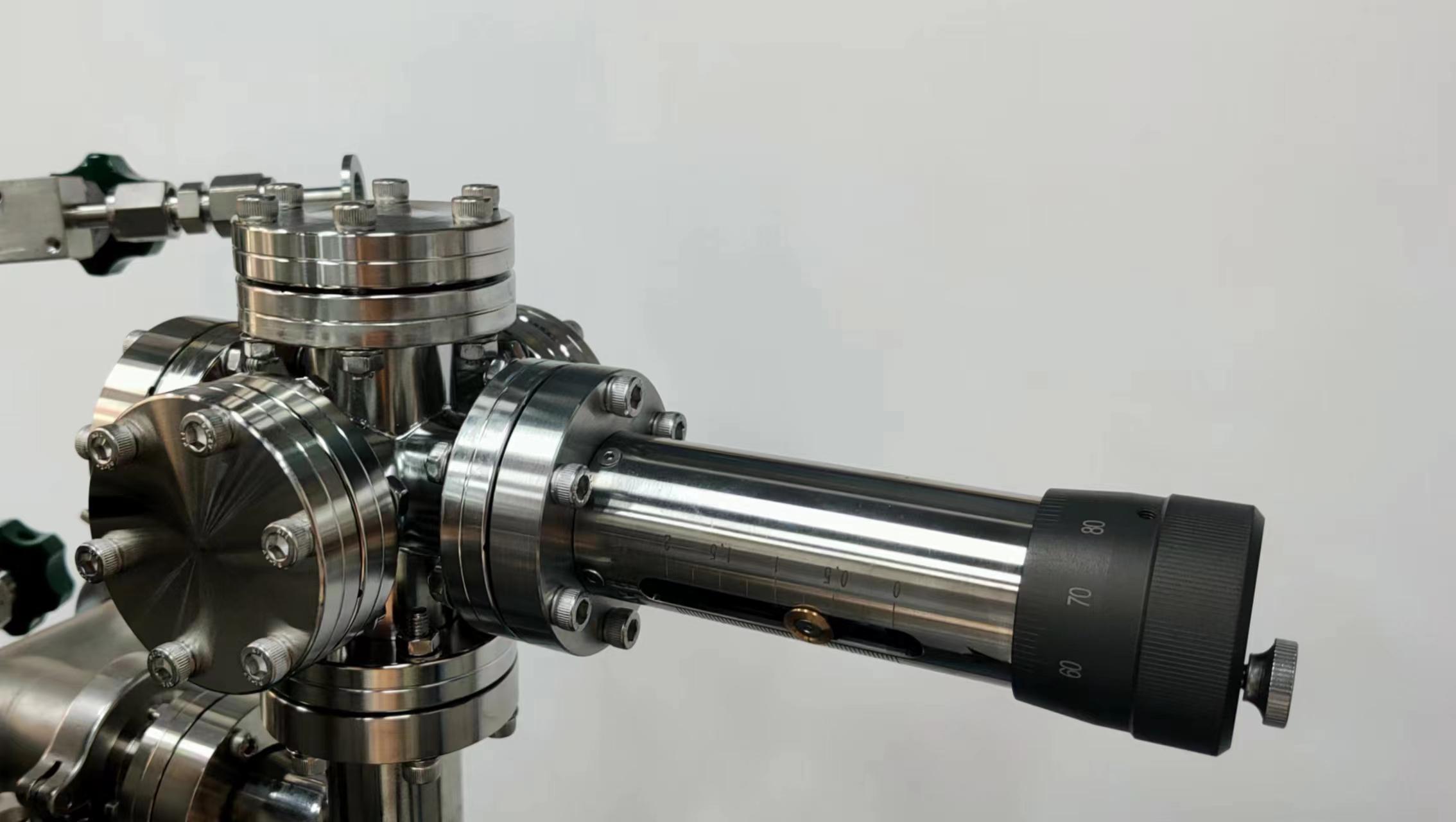}
}
\subfigure{
\label{fig:presure2}
\includegraphics[width=\hsize]{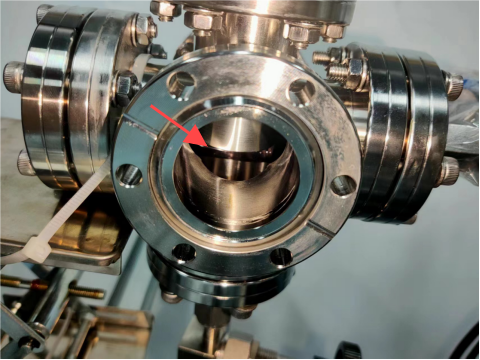}
}
\caption{(Color online) (top)Pressure conduction device for extracting and storing $\rm ^{37}Ar$ and (bottom) inner structure. (red arrow)The position of quartz ampoule.}
\label{fig:presure} 
\end{figure}

Following irradiation, the quartz ampoule was placed within a pressure-transfer apparatus, as indicated by the red arrow in Fig.~\ref{fig:presure}. The apparatus is shown in Fig.~\ref{fig:presure} was used for the precise recovery of all gases generated after irradiation. The process begins with the evacuation of the apparatus to achieve vacuum, thereby eliminating any extraneous atmospheric influences. The release of the trapped gas was achieved by applying pressure to the ampoule placed in the vacuum chamber via a pressure transfer apparatus with a maximum capacity of 100 N. The gas then diffuses and homogenizes within the system, allowing for controlled and quantified extraction of the gas according to experimental requirements, ensuring both the accuracy and integrity of the sample.

Based on the simulation results (Sect.~\ref{sec:simulation}), the yields and activities of nuclides such as $\rm ^{37}Ar$ and $\rm ^{39}Ar$ can be determined. Furthermore, the "burn-up" effect\cite{haxton}, which refers to the potential reaction of newly formed nuclides with neutrons to produce other particles, was evaluated. The calculations indicated that the "burn-up" effect was negligible under our experimental conditions.

\subsection{Thermal Neutron Irradiation Simulation} \label{sec:simulation}

$\rm ^{39}Ar$ is devastating for dark-matter search experiments. 
Consequently, the mitigation of background signals is essential. 
To precisely identify the nuclides generated during the production of $\rm ^{37}Ar$ and exclude those with extended half-lives that are difficult to eliminate once they are introduced into the detector, we performed a detailed simulation experiment. 
The purpose of this simulation was to emulate the actual irradiation conditions and to evaluate the probability of producing other potential nuclides. To achieve this, we established the following parameters for the simulation.

Based on the neutron flux in the reactor, a simulation was performed to ensure that the thermal neutron proportion was maintained at 24/25, and the remaining fraction consisted of epithermal neutrons. All neutrons were introduced randomly from the side to simulate the natural variability of the neutron incidence. To enhance the yield of isotopes other than $\rm ^{37}Ar$, particularly to amplify reactions with low probabilities during our simulation, we increased the proportion of isotopes other than $\rm ^{36}Ar$, which serves as the target nucleus for the production of $\rm ^{37}Ar$. When statistically analyzing the results, we adjusted the proportions to reflect the actual yields and effectively scaled the amplified ratios back. Tab.~\ref{tab:gasmat} presents the composition and mass fractions of all gases before the actual irradiation. This approach allows for a more accurate assessment of the production of nuclides during the irradiation process, ensuring that the sensitivity of the detector to dark matter signals is not compromised by the presence of long-lived background isotopes.

\begin{table}[!htb] 
\caption{The composition and mass fractions of all gases in the quartz ampoule before the irradiation.}
\label{tab:gasmat}
\begin{tabular*}{0.6\linewidth} {@{\extracolsep{\fill} }  c c}
\toprule
isotopes        & mass fractions (\%) \\
\midrule
$\rm ^{36}Ar$   & 99.935  \\
$\rm ^{38}Ar$   & 0.049  \\
$\rm ^{40}Ar$   & 0.004  \\
$\rm CH_{4}$    & 0.002  \\
$\rm CO/N_{2}$  & 0.002  \\
$\rm O_{2}$     & 0.003  \\
$\rm CO_{2}$    & 0.004  \\
$\rm H_{2}O$    & 0.001  \\
\bottomrule
\end{tabular*}
\end{table}

Our simulation, as indicated by the data presented in Table~\ref{tab:crosssection}, provided the cross-sections of the thermal neutron irradiation reaction and the half-lives of the selected argon isotopes \cite{haxton}. This table lists the cross-sections associated with the $(n,\gamma)$ process, with a particular emphasis on $\rm ^{37}Ar$, which uniquely possesses combined cross-sections for two distinct processes: $\sigma(n,p) + \sigma(n,\alpha) = (2040 \pm 340)$ barn. 
Table~\ref{tab:product} extends this analysis to encompass all potential nuclides and their respective yields generated at a simulated pressure of 0.1 bar within the ampoule. It is evident that in addition to $\rm ^{37}Ar$, the production probability of other nuclides is extremely low. 

\begin{table*}[!htb] 
\caption{The reaction cross-sections and half-life($\tau_{1/2}$) of argon isotopes with thermal neutrons\cite{haxton}.}
\label{tab:crosssection}
\begin{tabular*}{\linewidth} {@{\extracolsep{\fill} } l c c c c c c r}
\toprule
isotopes       & $\rm ^{36}Ar$ &	$\rm ^{37}Ar$ &	$\rm ^{38}Ar$ &	$\rm ^{39}Ar$ &	$\rm ^{40}Ar$   &	$\rm ^{41}Ar$ &	$\rm ^{42}Ar$ \\
\midrule
$\sigma$(barn) & $5.2 \pm 0.5$ &	$2040 \pm 340$&	$0.8 \pm 0.2$ &	$600 \pm 300$ &	$0.66 \pm 0.01$ &$0.5 \pm 0.1$    &	- \\
$\tau_{1/2}$   & stable        &	35.01 d       &	stable        & 269 yr        &	stable          &	1.83 h        &	33 yr \\
\bottomrule
\end{tabular*}
\end{table*}

\begin{table*}[!htb] 
\caption{The yield and decay information of possible generated nuclide with neutron irradiation at a pressure of 0.1 bar within the ampoule.}
\label{tab:product}
\begin{tabular*}{\linewidth} {@{\extracolsep{\fill} } l c c c c c}
\toprule
target        & generated               & yield        &	decay                           & half-life      & decay\\
nuclide       & nuclide                 & (s$^{-1}$)   &	mode                            & ($\tau_{1/2}$) &product\\ 

\midrule
\midrule

$\rm ^{28}Si$ & $\rm ^{29}Si$           &	-                       & stable                            &	-            &	-\\

\midrule

~             & $\rm ^{33}S$(stable)    &	$(3.95 \pm 0.20)$ E+5   & -                                 & -              & - \\
$\rm ^{36}Ar$ & $\rm ^{36}Cl$           &	$(1.65 \pm 0.09)$ E+5   &	$\rm \beta^{-}$/$\rm \beta^{+}$ & 3.01E+5 yr     &	$\rm ^{36}S$(stable)/$\rm ^{36}Ar$(stable) \\
~             & $\rm ^{37}Ar$           &	$(4.43 \pm 0.22)$ E+8   &	$\epsilon$                      & 35.01 d        &	$\rm ^{37}Cl$(stable) \\

\midrule

~             & $\rm ^{35}S$            &	-                       & 	$\rm \beta^{-}$                 & 87.35 d        &	$\rm ^{35}Cl$(stable) \\
$\rm ^{38}Ar$ & $\rm ^{38}Cl$           &	-                       &	$\rm \beta^{-}$                 & 37.24 min      &	$\rm ^{38}Ar$(stable) \\
~             & $\rm ^{39}Ar$           &	$(3.08 \pm 0.15)$ E+4   &	$\rm \beta^{-}$                 & 268 yr         &	$\rm ^{39}K$(stable) \\

\midrule

~             & $\rm ^{37}S$            &	-                       &	$\rm \beta^{-}$                 & 5.505 min      &	$\rm ^{37}Cl$(stable) \\
$\rm ^{40}Ar$ & $\rm ^{40}Cl$           &	-                       &	$\rm \beta^{-}$                 & 1.35 min       &	$\rm ^{40}Ar$(stable) \\
~             & $\rm ^{41}Ar$           &	$(2.02 \pm 0.10)$ E+3   &	$\rm \beta^{-}$                 & 109.61 min     &	$\rm ^{41}K$(stable) \\
\bottomrule
\end{tabular*}
\end{table*}

During the simulation, specific attention was directed towards two nuclides: $\rm ^{29}Si$ and $\rm ^{41}Ar$.
Although $\rm ^{29}Si$ exhibits a comparatively elevated yield, it is derived from the neutron irradiation of $\rm ^{28}Si$ present in quartz and is not expected to enter the gas source. In contrast, $\rm ^{41}Ar$, despite its certain yield, has a half-life of merely 109.61 min, indicating that it decays rapidly. 
Furthermore, the presence of $\rm ^{39}Ar$, if mixed uniformly with xenon within the detector, poses a challenge for removal, thus significantly increasing the background level of the detector. The simulation results substantiate our rationale for proceeding with subsequent experimental endeavors.

\subsection{The Gaseous Xenon Time Projection Chamber} 
\label{sec:gxe-tpc} 

Before injecting the $\rm ^{37}Ar$ calibration source into the ton-level detectors, it was injected into a GXe TPC to validate its performance. 
The detector was operated using gaseous xenon at room temperature.
Gas detectors represent a crucial subset of instruments utilized in particle and nuclear physics experiments. Time projection chambers that are made with gas as the medium have many applications in nuclear reactions and particle detection.
Xenon is chosen as the detection medium due to its pivotal role in dual-phase time projection chambers (LXe TPCs) used in dark matter and neutrino experiments such as PandaX-4T, XENONnT, LZ and so on.
The GXe TPCs have several notable advantages.
First, GXe TPC avoids the operational complexities associated with cryogenic and slow control systems.
Second, GXe TPCs feature a lower detection threshold and reduced background compared to LXe TPCs because the background is dominated by gamma rays and cosmic muons.
Additionally, argon and xenon, as members of the same group in the periodic table, exist in the gaseous phase at room temperature, enabling a uniform distribution within the detector. 
This uniformity is advantageous for measuring the activity of calibration sources and for facilitating the verification of activity estimations.
Although gaseous xenon emits fewer photons than liquid xenon, leading to reduced efficiency in detecting S1-S2 paired events, S2-only analysis can estimate the decay rate with high detection efficiency.

\begin{figure}[!htb] 
\includegraphics[width=1.0\linewidth]{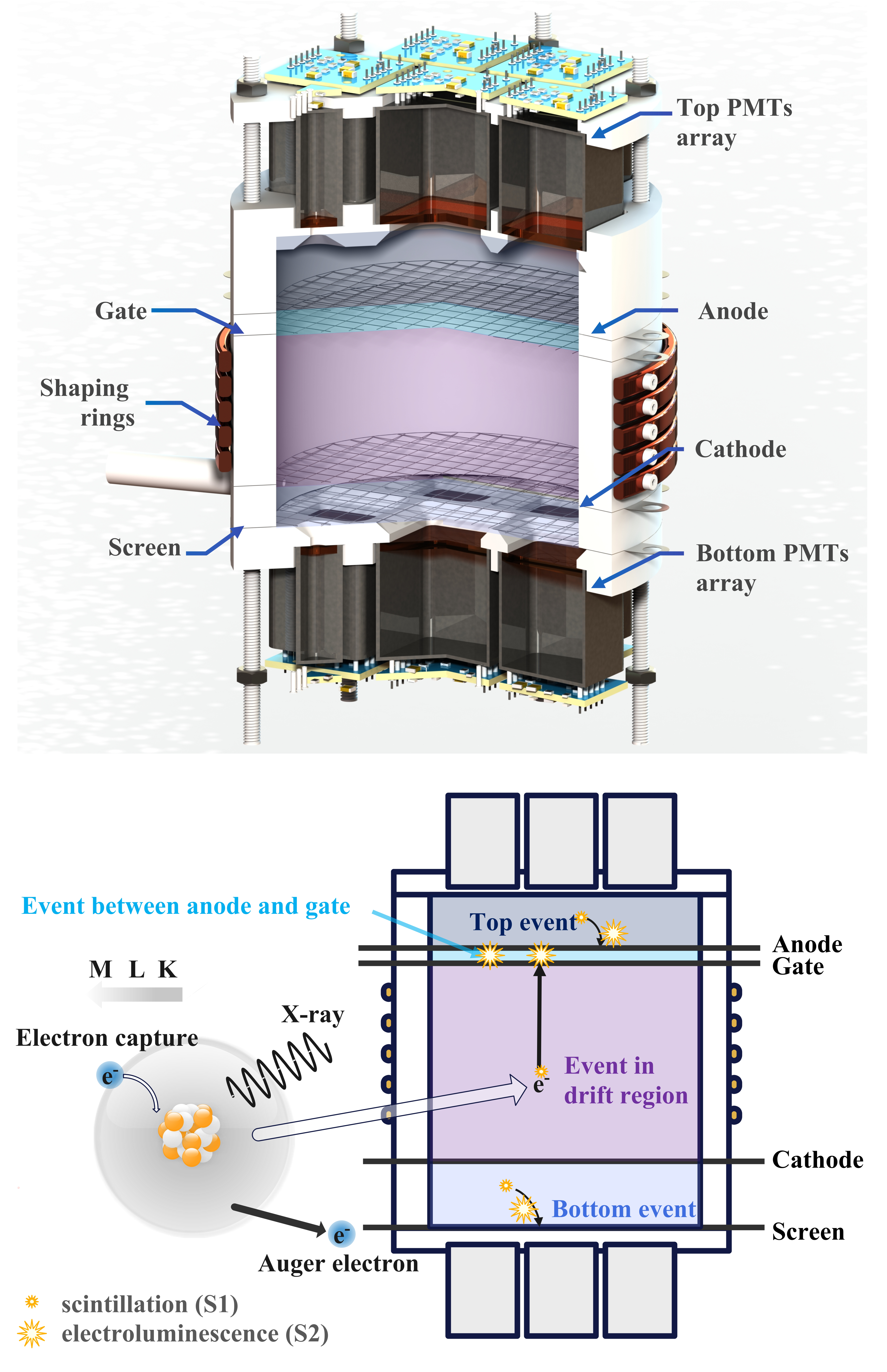}
\caption{(Color online) (Top) 3D rendered design of the RELICS demonstrator xenon time projection chamber. Blue lines and labels indicate detector components. This TPC is designed to operate in gaseous or liquid modes. (Bottom) The K, L, and M shell decays of $\rm ^{37}Ar$, along with a schematic diagram of events in the detector, are illustrated. Events occurring within the cathode and gate region (referred to as the drift region) are classified as regular events, while those in other regions are considered background signals.}
\label{fig:TPC}
\end{figure}

A schematic diagram of the GXe TPC used in this measurement is shown in the top panel of Fig.~\ref{fig:TPC}.
This TPC served as a prototype detector for the RELICS experiment~\cite{RELICS}. 
The TPC was mounted inside a double-walled cryostat to provide thermal insulation and structural support.  
It was equipped with 14 Hamamatsu R8520-406 PMTs, which were compactly placed on the top and bottom of the TPC and optimized for high VUV photon detection efficiency. 
These PMTs operate at a working voltage of $\rm -800\  V$.
Each array comprised seven PMTs in a regular hexagonal pattern positioned above and below the drift region.
The TPC walls are made of Teflon, which has excellent VUV reflectivity and enhanced light collection efficiency.
This arrangement provides relatively high light-collection efficiency and improves the spatial resolution of the detected events.

The bottom panel of Fig.~\ref{fig:TPC} shows the operational principle of the GXe TPC for detecting decay $\rm ^{37}Ar$.
$\rm ^{37}Ar$ decay produces scintillation and ionization electrons in the GXe. 
The scintillation photons were detected directly by the PMTs as the S1 signal. 
The ionization electrons drift under an electric field toward the proportional luminescence region, where they emit secondary scintillation light (S2).
The top and bottom arrays of photomultiplier tubes (PMTs) capture S1 and S2 signals, enabling precise event reconstruction, including its energy and three-dimensional positions.

The detector system integrates various subsystems, including cryogenic, gas purification, data acquisition, and recycling equipment subsystems.
The TPC operates at a pressure of approximately $170\ \rm kPa$, with gaseous xenon continuously circulated through a hot-getter system for purification. 
The purification process removes electronegative impurities, such as oxygen and water, which may absorb scintillation light and ionization electrons, reducing the detection and identification efficiency of $\rm ^{37}Ar$ decays. 
The electron drift region of the TPC is defined by a set of electrodes, including the anode, gate, cathode, and five shaping rings, which establish a uniform electric field for electron drift and convert electrons to proportional scintillation photons.
The anode was maintained at a voltage of $+1200\ \rm V$ to amplify the S2 signals, whereas the gate, cathode, and screen were set to $-1800\ \rm V$, $-2400\ \rm V$, and $-800\ \rm V$, respectively. 
This voltage configuration ensures stable operation, minimizes the risk of electrical breakdown, and provides the available conditions for the readout of single-electron S2 signals.
This measurement was based on the GXe TPC operation mode to evaluate the radioactivity of the source. 

\section{Measurement of $\rm ^{37}Ar$ radioactivity within the GXe TPC} 
\label{sec:mearsurement}

\subsection{ Injection of the $\rm ^{37}Ar$ source } 
\label{sec:injection}

\begin{figure}[!htb] 
    \centering
    \includegraphics[width=1\linewidth]{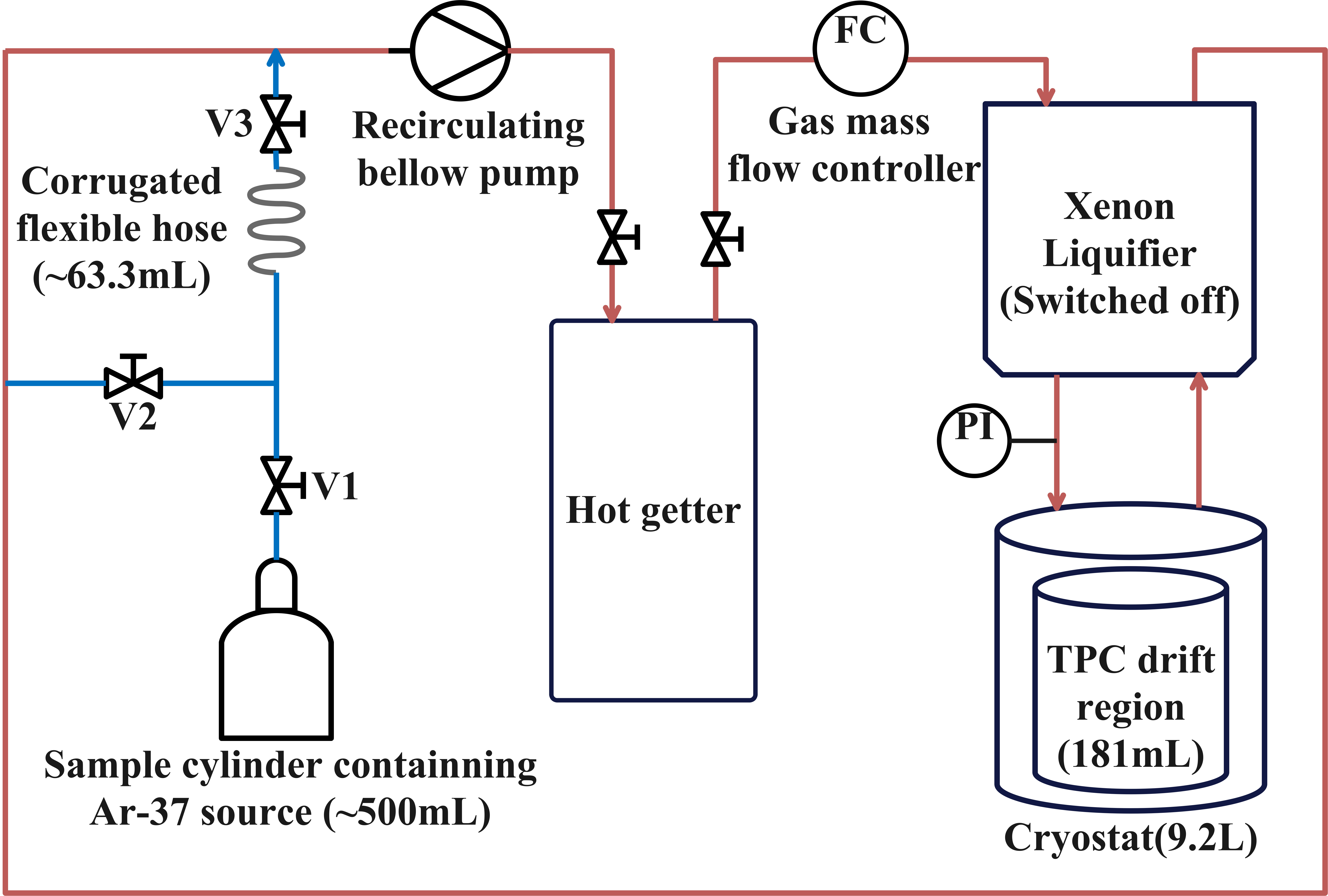}
    \caption{(Color online) Simplified diagram of the source injection, circulation and purification pipeline: the red line represents the recycling and purification route, while the blue line indicates the injection pathway.}
    \label{fig:PID}
\end{figure}

The $\rm ^{37}Ar$ source was stored in a Stainless Steel container with a volume of 500 mL.
A dedicated pipeline was developed to allow controlled introduction of a fixed portion of the $\rm ^{37}Ar$ source into the gaseous xenon detector system.
A simplified diagram illustrating the injection and gas recycling routes is shown in Fig. ~\ref{fig:PID}. This dosing system is designed to allow seamless calibration source injection during detector operation while minimizing the impact on xenon gas purity.
The activity of the injected source was calculated based on the volumetric relationships between the pipeline (including the cryostat containing the GXe TPC), storage container, and drift region of the TPC, assuming a uniform distribution of $\rm ^{37}Ar$. 
Detailed information regarding the volumes within the injection system is provided in Tab. ~\ref{tab:volume}.

\begin{table}[!htb] 
\centering
\caption{Component volumes of the injection source system.}
\label{tab:volume}
\begin{tabular*}{0.85\linewidth}{@{\extracolsep{\fill} } c c}
\toprule
Component            & Volume               \\ \midrule
50 cm long, VCR-1/2 pipeline         & $63.3\ \text{mL}$    \\
Source bottle        & $500\ \text{mL}$     \\
TPC drift region     & $181\ \text{mL}$     \\
Total system         & $28\ \text{L}$       \\ \bottomrule
\end{tabular*}
\end{table}

The $\rm ^{37}Ar$ source is introduced through multiple injections.
The circulation pipe enclosed by valves V1, V2, and V3 was defined as the dilution volume for the source injection.
Each injection was performed using several steps.
First, the dilution volume was pumped into a vacuum.
Then, $\rm ^{37}Ar$ was introduced to the dilution volume by opening V1. 
Consequently, 11\% of the total source was introduced into the dilution volume and injected into the circulation. 
The source was then uniformly distributed in the system with a total volume of $\simeq28$\,L. 
As the drift region of the TPC was only $\rm 181 mL$, another dilution factor of 0.6\% was introduced. As a result, only 0.07\% of the total radioactivity was measurable in GXe TPC.   

\subsection{ Data acquisition and signal processing} 
\label{sec:ar37-peaks}

To achieve a high detection efficiency of low-energy signals from the source $\rm ^{37}Ar$, all waveforms from the PMTs were digitized using CAEN V1725 digitizers, which employ a dynamic acquisition window (DPP-DAW) firmware for self-triggering readout. 
The digitized raw data were stored on a server, and subsequent event reconstruction and analysis were performed on dedicated analysis servers.
Data acquisition was carried out over an 8-hour period both before and after the injection of the $\rm ^{37}Ar$ source, allowing background subtraction. 
A software package was developed to process the data acquired from each PMT and to group them into peaks. 
A peak is defined as a waveform that features two or more PMT signals within $\rm \sim 300\,ns$. 
Scintillation and ionization signals from interactions with energy depositions in the GXe TPC, including decays in $\rm ^{37}Ar$, produce peaks in the data.

The area of a peak is proportional to the number of photons detected by the PMTs and is expressed in units of photon-electron~(PE), as calibrated by single-photon counting with an LED.
S1 peaks, induced by scintillation photons produced by direct excitation of the Xe atom or by recombination of electron and ion pairs from ionization, have a narrow distribution in time with a typical spread below $\rm \sim200\,ns$. 
S2 peaks, induced by the electroluminescence of the electrons drifting in GXe at a strong electric field~(notably between the Gate and Anode electrodes), have a wider distribution in time with a typical spread above $\rm \sim200\,ns$. 
The time spread of a peak is characterized by the leading time, defined as the time interval between the $\rm 0 \%$ and $\rm 50 \% $ percentiles of the waveform area. 
The relative peak area distribution on the PMT arrays depends on the light collection efficiency of each PMT,  and is used to reconstruct the position of an interaction. 
For the S2 peaks induced by interactions in the drift region, the horizontal distribution was reconstructed from the area distribution pattern on the top PMT array. 
S2 peaks can also be produced above the anode or below the cathode, because the detector is operated in the GXe mode.
The area fraction of the top~(AFT), which is the ratio of the area recorded by the top PMTs to the total area, is distinguishable for the S2 peaks produced in the drift region and below the cathode or above the anode.

\begin{figure}[!htb] 
    \centering
    \includegraphics[width=1.0\linewidth]{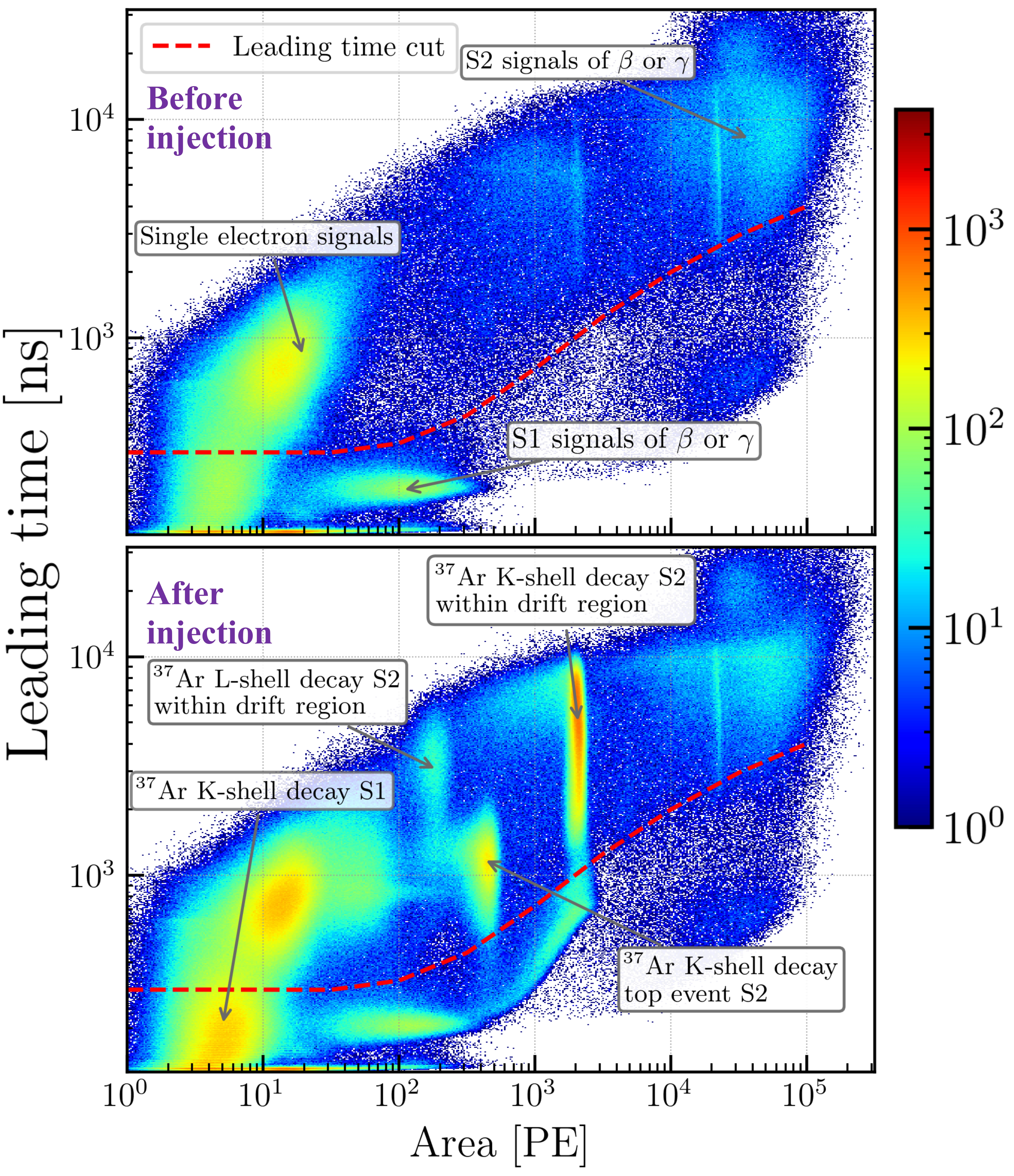}
    \caption{(Color online) Area and leading time of peaks distribution before and after source injection. (Top) Peaks collected before source injection. There may be a vestigial $\rm ^{37}Ar$ that was injected and decayed for several months in xenon. (bottom) Peaks collected after source injection. Significant $\rm ^{37}Ar$ signals arise at approximately $2000 \text{PE}$.
    }
    \label{fig:peaks_distribution}
\end{figure}

The distribution of the peaks in the area and leading-time space is shown in Fig. ~\ref{fig:peaks_distribution}. 
The peaks collected before and after injection of the $\rm ^{37}Ar$ source are shown in the top and bottom panels of Fig. ~\ref{fig:peaks_distribution}, respectively.
The pulses with a leading time above the dashed red line and an area greater than 100 PE are attributed to beta or gamma interactions within the drift region of the GXe TPC. 
pulses with an area of $\sim$ 20\,PE and a leading time of $\sim$ 700\,ns characterize S2 produced by single electrons drifting between the gate and anode. 
pulses with an area below 500\,PE and a leading time below the dashed red line correspond to S1s.

Some additional populations appear after the injection of the $\rm ^{37}Ar$ source: signals with an area of approximately 2000\,PE correspond to S2s from the K-shell $\rm ^{37}Ar$ electron capture events in the drift region. 
Pulses with an area of approximately 200\,PE correspond to S2s from the L-shell $\rm ^{37}Ar$ electron-capture events in the drift region. 
Pulses with an area below 10\,PE and leading time below the dashed red line correspond to S1s from the K-shell $\rm ^{37}Ar$ electron capture events. 
The identification is based on the known energy spectrum of $\rm ^{37}Ar$, in which the K-shell and L-shell electron capture lines are at approximately $\rm 2.8 keV$ and $\rm 0.27 keV$, respectively. Given the W-value of gaseous xenon (approximately $\rm 22.0\ eV$) and a single-electron gain of approximately 20 PE, the expected S2 area for K-shell events is approximately $\rm 2000\ PE$, and the L-shell contribution is approximately one-tenth that, which matches well with the observed populations.

In this study, we focus on the signals corresponding to $\rm ^{37}Ar$ K-shell decay events that occur within the region between the drift regions. 
The events detected outside this region were classified as background events.
To suppress these background events, it is necessary to know properties such as the light collection efficiency distribution and electron transport processes, which have not been thoroughly simulated, and the photon detection efficiency of PMTs remains insufficiently understood.
These factors introduce constraints in the accurate analysis of signals.
Consequently, a data-driven analysis approach was used to reduce the background and estimate the activity of the source $\rm ^{37}Ar$.
This method compensates for the lack of comprehensive detector simulations and allows evaluation of $\rm ^{37}Ar$ source activity.

\begin{figure}[!htb] 
    \centering
    \includegraphics[width=\linewidth]{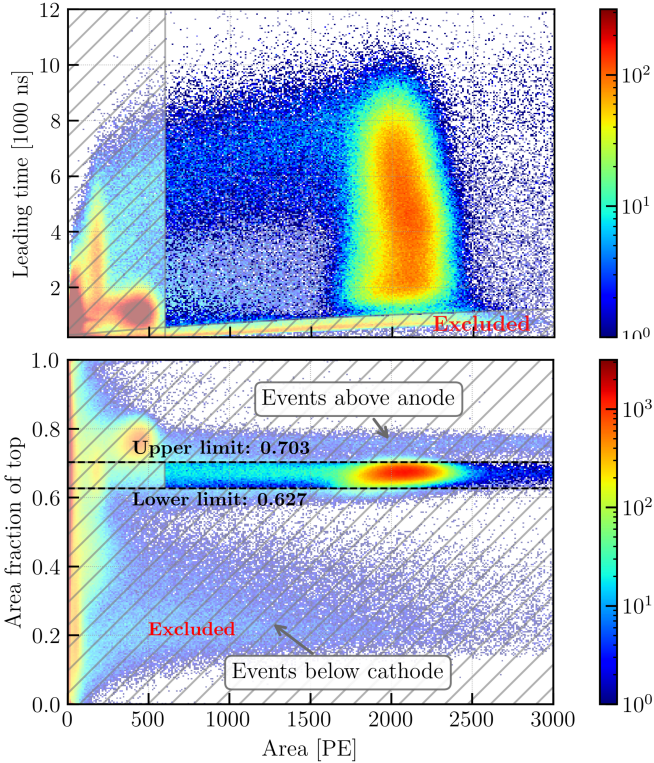}
    \caption{(Color online) Distribution of peaks' area and leading time after source injection. The acceptance probability for the area fraction of the top was $\rm 95.0\%$, whereas the acceptance probability for the leading time was $\rm 98.96\%$. The black dashed lines represent the selection thresholds for leading time and area fraction of top, with the shaded regions indicating the events excluded by these criteria.}
    \label{fig:preselected}
\end{figure}

The analysis focuses on the S2 signals, represented by the regions above the dashed red lines in Fig. ~\ref{fig:peaks_distribution}. 
Accurately determining the activity of $\rm ^{37}Ar$ requires meticulous data selection to minimize the impact of background noise.
As shown in Fig.~\ref{fig:preselected}, three different types of background noise are identified and removed.

First, events occurring between the anode and gate exhibited a positive correlation between the S2 area of these background signals and their leading time. 
These events  were located in the lower region of the distribution shown in the top panel of Fig. ~\ref{fig:preselected}, indicating a relationship between the event timing and background signal intensity.
Second, when the photomultiplier is set to $-800\ \rm V$ with a positive anode voltage, the ionized electrons generated by high-energy events can drift toward the anode under the influence of the electric field between the anode and top PMT array.
This drift results in peaks with a larger proportion of top PMTs.
Similarly, events that occur between the cathode and screen tend to produce a signal with a smaller area fraction of the top.
Furthermore, some pulses exhibited reduced light collection efficiency in specific regions, which appear on the left side of the distribution in the top panel of Fig. ~\ref{fig:preselected}. To correct for this bias, a crystal-ball model was employed to describe this phenomenon and fit the signal count.

These background events are effectively removed by selecting a waveform based on the area fraction of the top~(AFT) and the leading time. 
The distribution of the AFT for events at a fixed area in the drift region is described by a skew-Gaussian distribution to determine the acceptance of the cut. 
The cut boundary corresponding to the selection efficiencies of $\rm2.5 \%$ and $\rm 97.5 \%$, respectively, was determined to be $\rm \left( 0.627, 0.703 \right)$. Events occurring between the anode and gate have a similar area fraction to the top $i$th signal. 
They are characterized by shorter leading times compared to events occurring in the drift region, as the drift lengths for these ionization electrons are shorter. 
Peaks with leading times shorter than approximately $\rm 1030\ ns$ were excluded from this measurement, resulting in a selection efficiency of $\simeq 99 \% $.

\subsection{ $\rm ^{37}Ar$ K-shell activity estimate} 
\label{sec:act-ar37}

\begin{figure}[!htb] 
    \centering
    \includegraphics[width=\linewidth]{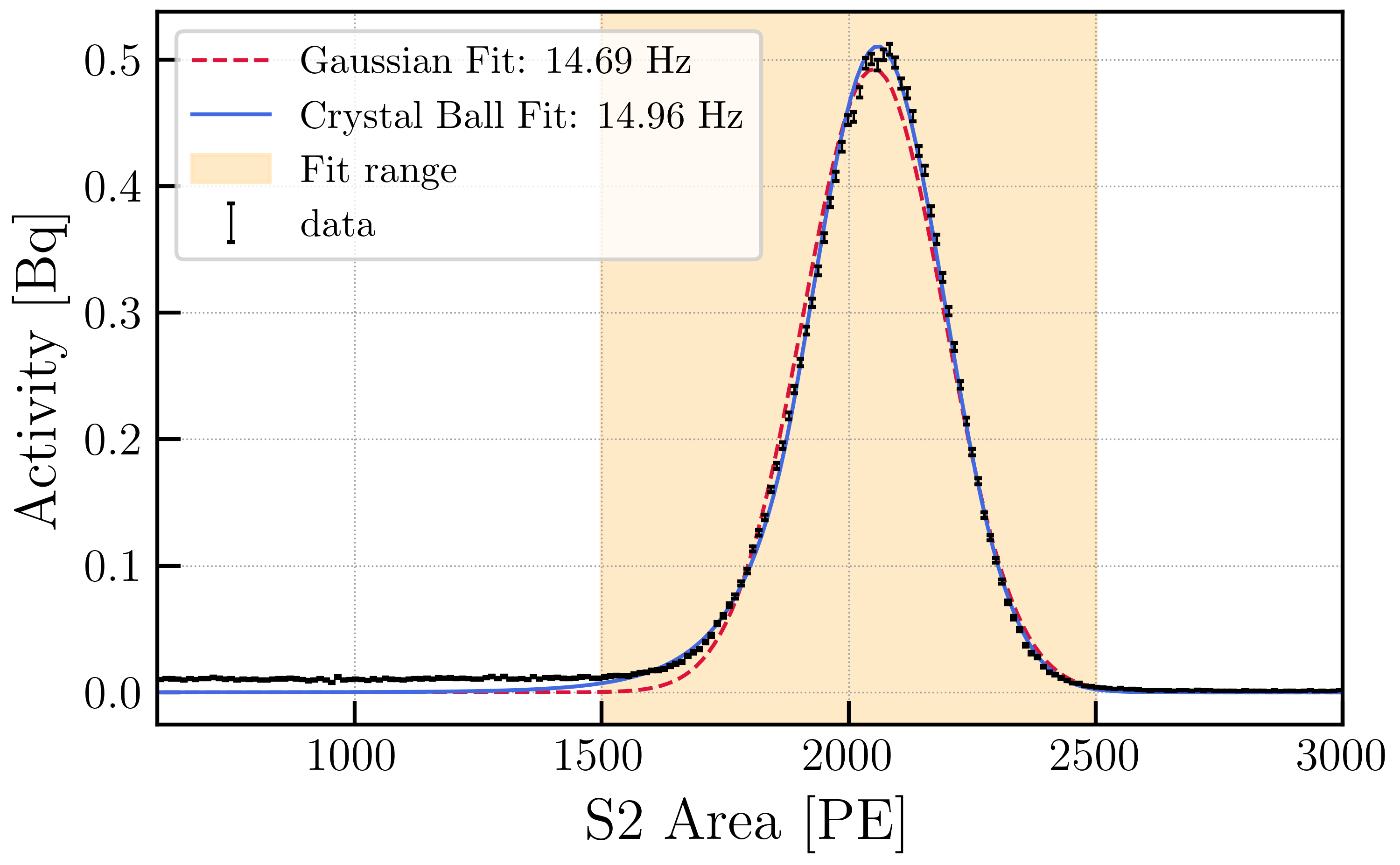}
    \caption{(Color Online) The final selected S2 spectrum of $^{37}$Ar K-shell decay, fitted using both Gaussian and Crystal-Ball distributions, with a resulting activity of approximately $\rm 14.96\ Bq$ by crystal-ball model. }
    \label{fig:area_fit}
\end{figure}

The magnitude distribution of the area was obtained after selecting the peaks.
The selected S2 spectrum from the $\rm ^{37}Ar$ K-shell decay was analyzed using Gaussian and Crystal Ball distributions to determine the event rates, as shown in Fig. ~\ref{fig:area_fit}.
The crystal band distribution was selected because it provides a more accurate representation of the spectrum, accounting for the effects of low photon detection efficiencies in certain regions of the projection chamber.
The Crystal Ball function combines a Gaussian core with a power-law tail, offering flexibility to model the asymmetric features observed in the spectrum. Mathematically, it is expressed as
$$
f(x; \alpha, n, \bar{x}, \sigma) =
\begin{cases}
A \exp\left(-\frac{(x - \bar{x})^2}{2\sigma^2}\right), & \text{for } \frac{x - \bar{x}}{\sigma} > -\alpha \\
B \left(C - \frac{x - \bar{x}}{\sigma}\right)^{-n}, & \text{for } \frac{x - \bar{x}}{\sigma} \leq -\alpha
\end{cases}
$$
where $\alpha$ determines the point at which the Gaussian transitions into the power-law tail, $n$ indicates the steepness of the power-law tail, and A and B are normalization constants that ensure continuity and smoothness at the transition point.

A fit using the Crystal Ball distribution yielded an observed activity of approximately $\rm 14.96\ Bq$. 
Considering that K-shell decays constitute 90.2\%  of all $\rm ^{37}Ar$ decays, and factoring in the selection efficiency of 94.0\% achieved through the area fraction of top (AFT) and leading time cuts, the total activity within the drift region is estimated at $\rm 17.646 \pm 0.025 (stat.) \pm 0.007 (sys.)\ Bq$. 
This activity level is well suited for calibrating liquid xenon dark matter detectors, such as PandaX-4T and XENONnT.

\section{Summary}\label{sec:sum}

In this study, we successfully synthesized the radioactive isotope $\rm ^{37}Ar$ using a reactor-derived thermal neutron source. 
With a half-life of 35.01 days, $\rm ^{37}Ar$ is particularly valuable for calibrating LXe TPCs in low-energy regions. The isotopes were produced by irradiating high-purity $\rm ^{36}Ar$ with thermal neutrons in a quartz ampoule. 
Geant4 simulations were used to predict the types and activities of the products, ensuring the minimal production of long-lived isotopes such as $\rm ^{39}Ar$.

The prepared $\rm ^{37}Ar$ source was injected into a GXe TPC for the preliminary measurements. 
Upon injection, a notable increase in the peak counts around 2000 PE was recorded, confirming the successful synthesis and deployment of the source. 
A data-driven analysis approach was applied to reduce the background noise and focus on the S2 signals of $\rm ^{37}Ar$ K-shell decay. 
The activity of the $\rm ^{37}Ar$ K-shell decay was measured to be approximately $\rm 14.96 Bq$. 
The conversion of $\rm ^{36}Ar$ to $\rm ^{37}Ar$ via neutron activation is a critical factor for determining the expected activity levels. 
An inaccurate estimation of the initial content of $\rm ^{36}Ar$ can lead to errors in calculating the decay rates and activities of $\rm ^{37}Ar$. 
This highlights the importance of precisely controlling the argon content during the preparation phase.
To mitigate this issue, a thorough review of the gas sealing process, particularly the impact of the temperature distribution during fusion sealing, could identify procedural errors that could contribute to underestimation.

In conclusion, this study successfully prepared and measured the activity of $\rm ^{37}Ar$, demonstrating its feasibility as a calibration source for low-energy dark-matter searches in LXe TPCs. 
These findings establish a solid foundation for future applications in detector calibration and dark-matter research.

\end{document}